\documentclass[preprint,5p,number,sort&compress,times]{elsarticle}

\usepackage{graphicx}
\usepackage{color}
\DeclareGraphicsExtensions{.eps,.png,.pdf,.jop}
\usepackage{subfigure}

\usepackage{citesort}
\include{aas_journals}

\usepackage{slashbox}
\usepackage{hyperref}

\usepackage{natbib}

\newcommand{\be}{\begin{equation}}
\newcommand{\ee}{\end{equation}}
\newcommand{\bea}{\begin{eqnarray}}
\newcommand{\eea}{\end{eqnarray}}
\newcommand{\bean}{\begin{eqnarray*}}
\newcommand{\eean}{\end{eqnarray*}}

\newcommand{\Mpc}{\mbox{\rm \,Mpc}}

\newcommand{\lcdm}{$\Lambda$CDM }
\begin{document}
\title{
    Reducing the Tension Between the BICEP2 and the Planck Measurements:\\
    A Complete Exploration of the Parameter Space
}

\author[naoc,ucas]{Yi-Chao~Li}
\ead{ycli@bao.ac.cn}

\author[naoc]{Feng-Quan~Wu}
\ead{wufq@bao.ac.cn}

\author[naoc]{You-Jun~Lu}
\ead{luyj@bao.ac.cn}

\author[naoc,chep]{Xue-Lei~Chen}
\ead{xuelei@cosmology.bao.ac.cn}

\address[naoc]{Key Laboratory for Computational Astrophysics,
    National Astronomical Observatories, Chinese Academy of Sciences,\\
20A Datun Road, Chaoyang District, Beijing 100012, China}
\address[ucas]{University of Chinese Academy of Sciences, Beijing 100049, China}
\address[chep]{Center of High Energy Physics, Peking University, Beijing 100871, China}


\date{\today}

\begin{abstract}
    A large inflationary tensor-to-scalar ratio 
    $r_\mathrm{0.002} = 0.20^{+0.07}_{-0.05}$ is 
    reported by the BICEP2 team based on their B-mode polarization detection, 
    which is outside of the $95\%$ confidence level of the 
    Planck best fit model.
    We explore several possible ways to reduce the tension between the 
    two by considering a model in which  $\alpha_\mathrm{s}$, $n_\mathrm{t}$, 
    $n_\mathrm{s}$ and the neutrino parameters $N_\mathrm{eff}$ and 
    $\Sigma m_\mathrm{\nu}$ are set as free parameters. Using the 
    Markov Chain Monte Carlo (MCMC) technique to survey the
    complete parameter space with and without 
    the BICEP2 data, we find that the resulting
    constraints on $r_\mathrm{0.002}$ are consistent with each other
    and the apparent tension seems to be relaxed. Further 
    detailed investigations 
    on those fittings suggest that $N_\mathrm{eff}$ probably plays the most 
    important role in reducing the tension. We also find that the results
    obtained from fitting without adopting the consistency relation do not
    deviate much from the consistency relation. 
    With available Planck, WMAP, BICEP2 and BAO 
    datasets all together, we obtain
    $r_{0.002} = 0.14_{-0.11}^{+0.05}$, 
    $n_\mathrm{t}  = 0.35_{-0.47}^{+0.28}$,
    $n_\mathrm{s}=0.98_{-0.02}^{+0.02}$, and
    $\alpha_\mathrm{s}=-0.0086_{-0.0189}^{+0.0148}$; 
    if the consistency relation is adopted, we get
    $r_{0.002} = 0.22_{-0.06}^{+0.05}$. 
\end{abstract}


\maketitle

\section{Introduction}
The BICEP2 experiment\cite{2014arXiv1403.3985B,2014arXiv1403.4302B},
a dedicated cosmic microwave background (CMB) polarization experiment,
has announced recently the detection of the B-mode polarization 
in CMB, based on an observation of about 380 square degrees 
low-foreground area of sky during 2010 to 2012 in the South Pole.
The detected B-mode power is in the multipole range $30<\ell<150$. 
Because the CMB lensing peaks at  $\ell\sim1000$,
the excess of B-mode power at these small $\ell\sim100$s 
can not be explained by the lensing contribution, which is too small.
It has been pointed out in that the foreground residual from Galactic dust
may contribute to B-mode power  \cite{2014arXiv1404.1899L, 2014arXiv1405.5857M,
2014arXiv1405.7351F}.  The BICEP2 team has examined possible 
systematic errors and potential foreground contaminations, and found that
the cross-correlations between frequency bands have little
changes in the observed amplitude, which imply that frequency-dependent 
foreground may not be the dominant contributor.  If the 
CMB polarization B-modes observed by BICEP2 is confirmed, it would
indicate the presence of 
tensor perturbations, i.e. gravitational waves in the early universe, 
and provide a strong evidence of the inflationary origin of the universe.

The inflation theory which has been developed since the 1980s 
solves a number of cosmological
conundrums, like the monopole, horizon, smoothness, and entropy problems
\cite{1980PhLB...91...99S, 1981PhRvD..23..347G, 1982PhRvL..48.1220A, 
1982PhLB..108..389L}.
The quantum fluctuations streched by the inflationary
expansion, give rise the scalar and tensor primordial power spectrum. 
Considering the \lcdm model and assuming the scalar perturbation are purely
adiabatic, it is convenient to expand the scalar and tensor power spectrum
as
\begin{eqnarray}
    P_\mathrm{\zeta} \left( k \right) &\equiv& A_\mathrm{s} \left( \frac{k}{k_0} \right)^{n_\mathrm{s} - 1 + \frac{1}{2} \alpha_\mathrm{s} \ln \frac{k}{k_0} }\mathrm{,} \;\;
    \label{parametrizationP}\\
    P_\mathrm{t} \left( k \right) &\equiv& A_\mathrm{t} \left( \frac{k}{k_0} \right)^{n_\mathrm{t}  }\mathrm{,}\;\;
\end{eqnarray}
where $k_0$ is the pivot scale, it is usually chosen to be 
$0.05$\Mpc$^{-1}$, roughly in the 
middle of the logarithmic range of scales probed by WMAP and Planck experiments;
$A_\mathrm{s}$, $n_\mathrm{s}$ are the amplitude and spectral index for 
the scalar power spectrum respectively, 
while $A_\mathrm{t}$, $n_\mathrm{t}$ are for the tensor power spectrum 
respectively; $\alpha_\mathrm{s}$ denotes the running of the scalar spectrum 
tilt\cite{1995PhRvD..52.1739K} with 
$\alpha_\mathrm{s} = \frac{d \, n_\mathrm{s}}{d \, {\rm ln }  \, k}$.
An important parameter, the tensor-to-scalar ratio, which indicates the ratio
of the tensor power to the scalar power, is defined as 
\begin{equation}
    r = \frac{P_\zeta(k)}{P_t(k)}\mathrm{,}
\end{equation}
$r$ can be scale dependent, and the single field slow-roll inflation implies a
tensor-to-scalar ratio of $r_{0.05} = -8n_\mathrm{t}$, in which the 
subscript $0.05$ indicates the particular pivot 
scale of $k=0.05$\Mpc$^{-1}$. This relation is referred
as the {\it consistency relation}. 

The BICEP2 team reported their
measured value of tensor-to-scalar ratio, at scale $k=0.002$\Mpc$^{-1}$, as
$r_{0.002}=0.20^{+0.07}_{-0.05}$, based on the lensed-$\Lambda$CDM+tensor 
model. The result is derived from importance sampling of the Planck MCMC chains 
using the direct likelihood method. 
The unexpected large tensor-to-scalar ratio generated a lot of 
interests 
\cite{2014arXiv1403.5253H, 2014arXiv1403.5549C,
2014arXiv1403.4585M, 2014arXiv1403.5716G, 2014arXiv1403.7623X, 2014arXiv1404.2560C,
2014arXiv1403.3919Z, 2010PhRvD..82l3008Z,
2014arXiv1408.3966B, 2014arXiv1408.3192N, 2014arXiv1408.1285E, 2014arXiv1408.0046D,
2014arXiv1407.8105A, 2014arXiv1407.7692M}. 
There appears a tension between the value of $r_\mathrm{0.002}$
measured by the BICEP2 team and that by other CMB experiments, at least in the 
simplest lensed $\Lambda$CDM+tensors model.

Previous CMB observations with the Planck satellite, the WMAP satellite and 
other CMB experiments yielded a limit of much smaller 
tensor-to-scalar ratio $r<0.11$ (at $95\%$ C.L.)\cite{2013arXiv1303.5076P}.
Some mechanisms have been proposed to alleviate this tension 
\cite{2014arXiv1403.4596C}, 
by (a) adjusting the running of the scalar power spectrum tilt; 
(b) considering the blue tilt tensor power spectrum; and (c) including the
effect of the neutrinos.

\paragraph{The running of the scalar power spectrum} 
The BICEP2 team \cite{2014arXiv1403.3985B} pointed out that 
a simple way to relax this tension is to take the running of spectrum 
index into account, but large $|\alpha_\mathrm{s}|$ leads to an unacceptably
small value of e-folds number for slow roll inflation\cite{2006JCAP...09..010E}.

\paragraph{The blue tilt tensor power spectrum} 
There are wide spread interests in the tensor power spectrum 
index\cite{2014arXiv1403.5817W,2014arXiv1403.6099A}, since it 
is an important source of information for distinguishing inflation 
models \cite{2014arXiv1403.5922A,2014arXiv1403.5163G,2014arXiv1403.4927B}. 
Recently, \citet{2014arXiv1403.5732G} reports a blue tensor power spectrum tilt
$n_\mathrm{t}\sim 2$ using the B-mode measurements.
It is also possible to solve the tension by including $n_\mathrm{t}$ as 
a free parameter. 
\citet{2014arXiv1403.6462W} studies the effect of $n_\mathrm{t}$,
by including $\Omega_\mathrm{c}h^2$, $\Omega_\mathrm{b}h^2$, $\tau$, 
$\theta_\mathrm{MC}$, $A_\mathrm{s}$, $n_\mathrm{s}$ and $n_\mathrm{t}$
as free parameters in the global fitting, and finds that the apparent tension
is alleviated.

\paragraph{The effect of the neutrinos} 
Besides directly adjusting the spectrum itself, considering 
the effect of neutrinos may also suppress the scalar 
power spectrum. The effective number of neutrinos $N_{\mathrm{eff}}$ affects the 
density of the radiation in the universe, 
which change the expansion rate before 
recombination, and the age of the universe at recombination. The diffusion
length scales and sound horizon, which are all related with the age, affect the
power in its damping tail.\cite{2004PhRvD..69h3002B,2013PhRvD..87h3008H}.
Very massive neutrinos could suppress the structure formation at small 
scales \cite{2014arXiv1403.7028Z,2014arXiv1404.1794A}, though as there 
are tight limits on the mass of the three active neutrinos, such a neutrino
must be a sterile one.
It is reported that considering the effect of the neutrinos can reduce
the tension\cite{2014arXiv1403.7028Z, 2014arXiv1403.8049D,2014arXiv1408.0481Z}.

In this paper, we explore the best way to solve the tension, 
through the global fitting, by considering $\alpha_\mathrm{s}$, $n_\mathrm{t}$ 
as well as the neutrino parameters as free parameters. In the lensed $\Lambda$CDM 
model, the fitting is performed with the Planck CMB temperature 
data \cite{2013arXiv1303.5076P} and the WMAP 9 year CMB polarization 
data \cite{2013ApJS..208...19H, 2013ApJS..208...20B}, with/out the 
newly published BICEP2 CMB B-mode polarization data.
In order to have good constraints, the BAO data from the 
SDSS DR9 \cite{2014MNRAS.441...24A}, SDSS DR7 \cite{2012MNRAS.427.2132P}, 
6dF \cite{2011MNRAS.416.3017B} are also included. 
We derive constraints using the publicly available code 
COSMOMC \cite{2002PhRvD..66j3511L},
which implements a Metropolis-Hastings algorithm to perform 
a MCMC simulation in order to fit the cosmological
parameters. This method also provides reliable error estimates
on the measured variables.

The outline of this paper is as follows. 
In Sect.~\ref{sec:th_analysis} we
firstly check the sensitive scale for some interesting parameters, which could 
solve the tension; 
In Sect.~\ref{sec:fitting} we introduce our global fitting method and present the 
results; The contributions of the interesting parameters are discussed in 
In Sect.~\ref{sec:discussion}, and our conclusions are given in 
Sect.~\ref{sec:conclusion}.


\section{The sensitive scale for parameters}\label{sec:th_analysis}

\begin{figure}[htbp]
\begin{center}
\includegraphics[width=3.5in]{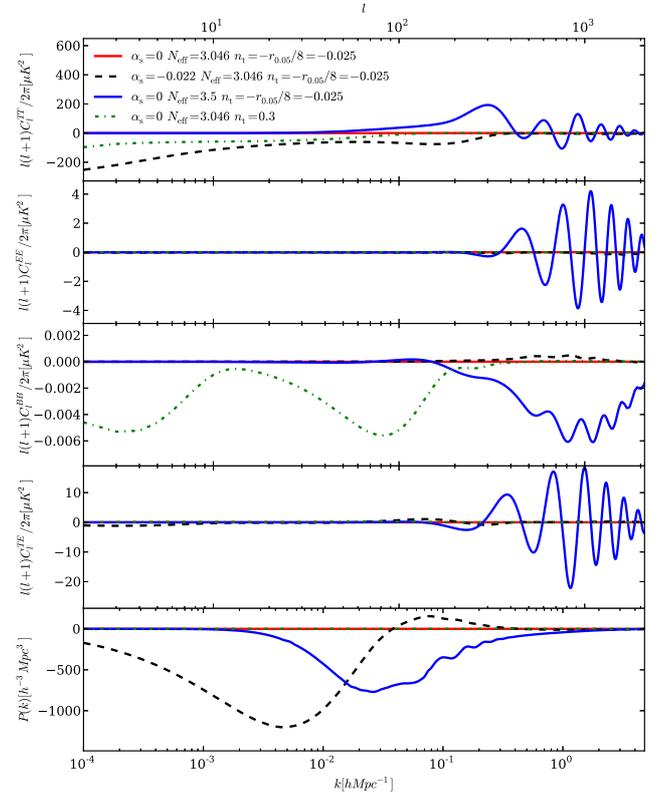}
\caption{
The difference in power spectra between a model with the fiducial parameters set.
From the top to the bottom are for the CMB TT, EE, BB, TE angular power spectra,  
and the matter power spectrum. The red solid line (actually the $x$-axis) 
indicates the fiducial case. The difference induced by variation in 
$\alpha_\mathrm{s}$(black dashed), $N_{\mathrm{eff}}$(blue solid), 
and $n_\mathrm{t}$(green dash dot) are plotted. 
}
\label{fig:power_test}
\end{center}
\end{figure}

The interesting parameters  $\alpha_\mathrm{s}$, $N_\mathrm{eff}$
and $n_\mathrm{t}$ are sensitive to different scales of the power spectrum.
Using the CAMB code, we can find out the sensitive scales of 
each parameter. For comparison, a baseline model is set with 
$\alpha_\mathrm{s}=0$, $N_{\mathrm{eff}} = 3.046$ and 
$n_\mathrm{t} = -r_{0.05}/8$, following the consistence relation. 
The fiducial values of the parameters are based on the result of Planck, 
except $N_\mathrm{eff}$, which comes from the Standard Model.
The residuals comparing with the fiducial power spectrum
are shown in Figure~\ref{fig:power_test}, in which the fiducial case is shown as the
red solid line.

The running of the power spectrum index is included by setting 
$\alpha_\mathrm{s}=-0.022$, and the residuals are shown as black dashed line in 
Figure~\ref{fig:power_test}. The result indicates that, $\alpha_\mathrm{s}$ is 
most sensitive to the scale with $\ell<200$ in the CMB angular spectrum and 
$\rm{k}<0.1\rm{hMpc}^{-1}$ in the matter power spectrum. Within such scales, the 
negative $\alpha_\mathrm{s}$ can reduce the TT and the matter power 
spectrum, which is expected for solving the problem in tensor-to-scalar
ratio. As reported in Ref.\cite{2013arXiv1303.5076P}, when 
$\alpha_\mathrm{s} = -0.022\pm0.01~(68\%)$, the constraints relax to 
$r_{0.002}<0.26$, which indicates a possible way to relax the tension. 

The parameter $N_{\mathrm{eff}}$ has great effect on smaller scales of the 
power spectrum. The Standard Model value is  
$N_{\mathrm{eff}} = 3.046$ \cite{2005NuPhB.729..221M}, we plot the difference   
result for the case $N_{\mathrm{eff}} = 3.5$  as shown by the blue solid line 
in Figure~\ref{fig:power_test}. $N_{\mathrm{eff}}$ affects the peaks of BAO, both on
the position and the amplitude \cite{2004PhRvD..69h3002B,2013PhRvD..87h3008H}, which 
is also clearly shown in our figure. 
A large $N_\mathrm{eff}$ causes the suppression on the small scales of the scaler
power spectrum. With a large $n_\mathrm{s}$,
the scalar power spectrum increases at the scales both larger and smaller than
the pivot scale of $k=0.05\mathrm{Mpc}^{-1}$. The increased power compensates the 
suppression at small scales and also reduces tensor-to-scaler ratio at large
scales, which can help reduce the tension of $r_\mathrm{0.002}$.

The green dashed-dot line in Figure~\ref{fig:power_test} shows the difference
between a model with $n_\mathrm{t}$ following the 
consistency relation and a model where this relation is broken,
with $n_\mathrm{t}=0.3$. It is shown that the variation of $n_\mathrm{t}$ 
mainly affects the large scales of the BB power spectrum. 

From the above discussions we see 
how each parameter affects the CMB and matter power spectra differently, and 
how it could help to alleviate the tension in the tensor-to-scalar ratio.
However, there are still degeneracy and correlation between the effects of 
various parameters, and the constraints also depends on the priors, so the actual 
result is more complicated. We perform a global fitting with complete parameter 
space, and flat prior, to explore the best way to solve the problem. 


\section{The global fitting}\label{sec:fitting}

\begin{table*}[htbp]
\centering
\caption{
    Cosmological parameters used in our analysis. For each of them, we list
    the symbol, prior range and the summary definition. Flat priors are assumed for 
    all parameters.
}
\begin{tabular}{ccl} \hline\hline
Parameter & Prior range & Definition\\
\hline
$\Omega_{\mathrm{c}} h^2$ & [$ 0.001$, $ 0.990$] & physical CDM matter density
\\
$\Omega_{\mathrm{b}} h^2$ & [$ 0.005$, $ 0.100$] & physical baryon density
\\
$\tau$ & [$ 0.010$, $ 0.800$] & Thomson scattering optical depth due to reionization 
\\
$100\theta_{\mathrm{MC}}$ & [$ 0.500$, $10.000$] & 100 times the ratio of sound horizon to angular-diameter distance to CMB last-scattering surface
\\
$\ln(10^{10} A_\mathrm{s})$ & [$ 2.700$, $ 4.000$] & Log power of the primordial curvature perturbations ($k_0 = 0.05 Mpc^{-1}$ )
\\
$n_\mathrm{s}$ & [$ 0.800$, $ 1.140$] & Scalar spectrum power-law index ($k_0 = 0.05 Mpc^{-1}$ )
\\
\hline
$\Sigma m_\nu\,[\mathrm{eV}]$ & [$ 0.000$, $ 5.000$] & sum of physical masses of standard neutrinos
\\
$N_{\mathrm{eff}}$ & [$ 3.046$, $ 8.000$] & effective number of neutrinos 
\\
$A_{\mathrm{L}}$ & [$ 0.000$, $ 5.000$] & lensing potential scaled by sqrt($A_{lens}$)
\\
$n_\mathrm{t}$ & [$-3.000$, $ 4.000$] & Tensor spectrum power-law index ($k_0 = 0.05 Mpc^{-1}$ ) \\
$\alpha_\mathrm{s}$ & [$-0.200$, $ 0.170$] & Running of the spectral index, $d n_\mathrm{s} / d\ln k$    
\\
$r_{0.05}$ & [$ 0.000$, $ 1.000$] & ratio of tensor to scalar primordial power at pivot scale $0.05 Mpc^{-1}$
\\
\hline
$m_{\nu,\,\mathrm{sterile}}^{\mathrm{eff}}$ & [$ 0.000$, $ 3.000$] & effective mass of sterile neutrino (eV) 
\\
\hline\hline
\end{tabular}

\label{tab:parameters}
\end{table*}

\begin{table*}[htbp]
\centering
\caption{
    The results of the global fitting with different datasets. For each
    of the fitting, we consider both imposing and not imposing the inflation consistency
    relation. Without the consistency relation, $n_\mathrm{t}$ is constrained by
    MCMC as a free parameter. The error are the $68\%$ marginalized limits.
    The columns with label ``Planck + WP" indicate the results obtained with only the Planck
    and WMAP datasets; the columns with label ``+ BICEP" indicate the result
    of ``Planck + WP + BICEP"; while ``+ BAO" indicate the results of 
    ``Planck + WP + BICEP + BAO".
}
\begin{tabular}{c||c|c|c||c|c|c} \hline\hline
& \multicolumn{3}{c||}{$n_\mathrm{t}$ free}& \multicolumn{3}{c}{$n_\mathrm{t} = -r_{0.05}/8$}\\
\hline
Parameter                       & Planck + WP                         & + BICEP                             & + BAO                               & Planck + WP                         & + BICEP                             & + BAO                                \\
\hline                                                                                                                                                                                                                                                                
$n_\mathrm{s}$                  & $  0.9929\pm_{  0.0396}^{  0.0334}$ & $  1.0062\pm_{  0.0357}^{  0.0332}$ & $  0.9804\pm_{  0.0219}^{  0.0177}$ & $  0.9995\pm_{  0.0422}^{  0.0374}$ & $  1.0159\pm_{  0.0389}^{  0.0363}$ & $  0.9815\pm_{  0.0225}^{  0.0182}$  \\
$r_{0.002}$                     & $<  0.1611                        $ & $  0.1597\pm_{  0.1247}^{  0.0638}$ & $  0.1359\pm_{  0.1056}^{  0.0526}$ & $<  0.2409$                         & $  0.2404\pm_{  0.0737}^{  0.0503}$ & $  0.2223\pm_{  0.0643}^{  0.0465}$  \\                                          
$n_\mathrm{t}$                  & $  0.2418\pm_{  0.4591}^{  0.2089}$ & $  0.2921\pm_{  0.4662}^{  0.2252}$ & $  0.3486\pm_{  0.4707}^{  0.2765}$ & -                                   & -                                   & -                                    \\
$\alpha_\mathrm{s}$             & $ -0.0054\pm_{  0.0217}^{  0.0162}$ & $ -0.0011\pm_{  0.0241}^{  0.0187}$ & $ -0.0086\pm_{  0.0189}^{  0.0148}$ & $ -0.0065\pm_{  0.0215}^{  0.0174}$ & $ -0.0017\pm_{  0.0227}^{  0.0185}$ & $ -0.0133\pm_{  0.0162}^{  0.0129}$  \\
$N_{\mathrm{eff}}$              & $  3.6572\pm_{  0.9201}^{  0.6088}$ & $  3.9273\pm_{  0.9277}^{  0.6641}$ & $  3.4284\pm_{  0.5834}^{  0.3902}$ & $  3.7774\pm_{  0.9911}^{  0.6788}$ & $  4.1128\pm_{  1.0179}^{  0.7262}$ & $  3.4220\pm_{  0.5813}^{  0.4021}$  \\
$\Sigma m_\nu\,[\mathrm{eV}]$   & $<  0.2850                        $ & $  < 0.2698$                        & $<  0.3698$                         & $<  0.3374$                         & $  <  0.3136$                       & $<  0.4115$                          \\
$\Omega_{\mathrm{m}}$           & $  0.2820\pm_{  0.0426}^{  0.0358}$ & $  0.2658\pm_{  0.0368}^{  0.0295}$ & $  0.2993\pm_{  0.0134}^{  0.0119}$ & $  0.2766\pm_{  0.0468}^{  0.0357}$ & $  0.2566\pm_{  0.0359}^{  0.0298}$ & $  0.2983\pm_{  0.0122}^{  0.0121}$  \\
$\Omega_\Lambda$                & $  0.7180\pm_{  0.0358}^{  0.0426}$ & $  0.7342\pm_{  0.0295}^{  0.0368}$ & $  0.7007\pm_{  0.0119}^{  0.0134}$ & $  0.7234\pm_{  0.0357}^{  0.0468}$ & $  0.7434\pm_{  0.0299}^{  0.0359}$ & $  0.7017\pm_{  0.0121}^{  0.0122}$  \\
$\sigma_8$                      & $  0.7913\pm_{  0.0321}^{  0.0485}$ & $  0.7985\pm_{  0.0320}^{  0.0455}$ & $  0.7770\pm_{  0.0360}^{  0.0514}$ & $  0.7842\pm_{  0.0352}^{  0.0543}$ & $  0.7942\pm_{  0.0326}^{  0.0487}$ & $  0.7689\pm_{  0.0397}^{  0.0525}$  \\
$H_0$                           & $ 72.8437\pm_{  7.2104}^{  5.3682}$ & $ 75.4450\pm_{  6.7827}^{  5.4714}$ & $ 69.9689\pm_{  2.6099}^{  2.1843}$ & $ 73.9136\pm_{  7.7843}^{  6.1711}$ & $ 77.1967\pm_{  7.5416}^{  5.8544}$ & $ 70.0222\pm_{  2.6669}^{  2.2480}$  \\
$100\theta_{\mathrm{MC}}$       & $  1.0412\pm_{  0.0010}^{  0.0009}$ & $  1.0411\pm_{  0.0010}^{  0.0010}$ & $  1.0412\pm_{  0.0009}^{  0.0009}$ & $  1.0412\pm_{  0.0010}^{  0.0009}$ & $  1.0411\pm_{  0.0010}^{  0.0009}$ & $  1.0413\pm_{  0.0009}^{  0.0009}$  \\
$A_{\mathrm{L}}$                & $  1.1454\pm_{  0.1367}^{  0.1003}$ & $  1.1871\pm_{  0.1245}^{  0.0996}$ & $  1.1277\pm_{  0.1036}^{  0.0799}$ & $  1.1756\pm_{  0.1572}^{  0.1095}$ & $  1.2258\pm_{  0.1436}^{  0.1110}$ & $  1.1447\pm_{  0.1086}^{  0.0856}$  \\
\hline\hline
\end{tabular}

\label{tab:result}
\end{table*}

We use the CosmoMC code \cite{2002PhRvD..66j3511L}  
to explore the parameter space and 
obtain limits on cosmological parameters. In our MCMC simulations,
about 500000 samples are collected with 200 chains. The first $1/3$ of 
the samples is used for burning and not used for the final analysis. 

In addition to the BICEP2 data \cite{2014arXiv1403.3985B}, we use the
Planck CMB temperature data \cite{2013arXiv1303.5076P},
the WMAP 9 year CMB polarization data \cite{2013ApJS..208...19H, 2013ApJS..208...20B},
and the BAO data from the SDSS DR9 \cite{2014MNRAS.441...24A}, 
SDSS DR7 \cite{2012MNRAS.427.2132P} and 6dF \cite{2011MNRAS.416.3017B} in our 
cosmological parameter fitting. For clarity, we use the following labels to denote 
the different datasets,

\begin{itemize}
    \item\textbf{Planck+WP:} 
        The Planck high-$\ell$, low-$\ell$ temperature power 
        data\cite{2013arXiv1303.5076P}, and the WMAP9 polarization power 
        data\cite{2013ApJS..208...19H, 2013ApJS..208...20B} are adopted in the
        fitting;
    \item\textbf{Planck+WP+BICEP:}
        Beside the Planck and WMAP datasets, the BICEP2
        data\cite{2014arXiv1403.3985B,2014arXiv1403.4302B} is also included;
    \item\textbf{Planck+WP+BICEP+BAO:}  
        Beside the CMB measurement data, the BAO data from 
        SDSS DR9\cite{2014MNRAS.441...24A}, 
        SDSS DR7\cite{2012MNRAS.427.2132P}, 
        and 6dF\cite{2011MNRAS.416.3017B} are also include in the fitting.
\end{itemize}

The definition and prior range of some important parameters are listed
in Table~\ref{tab:parameters}. For most of the parameters, the 
flat priors are used as in the Planck analysis\cite{2013arXiv1303.5076P}. 
Beside the 6 parameters characterizing the simplest inflationary $\Lambda$CDM
model, $\Omega_\mathrm{c}h^2$, $\Omega_\mathrm{b}h^2$, $\tau$, $\theta_\mathrm{MC}$,
$A_\mathrm{s}$ and $n_\mathrm{s}$,
we also include the running of scalar power spectrum index $\alpha_\mathrm{s}$ 
and the parameters related with the neutrinos, such as the effective number of 
neutrinos $N_{\mathrm{eff}}$ and the sum of physical masses of standard neutrinos 
$\Sigma \mathrm{m_{\nu}}$. Because the evolution of sterile neutrinos is significantly
different, it is explored as an extra case, by including 
one more parameter, $m_{\nu,\,\mathrm{sterile}}^{\mathrm{eff}}$, the effective
mass of the sterile neutrinos. When we 
ignore the single field slow-roll consistency relation,
$n_\mathrm{t}$ is set to a fixed value, which can be positive or negative, 
allowing for both red and blue tilt. For comparison, we also run a set of MCMC chains, 
with $n_\mathrm{t}$ following the consistency relation.


\section{Results and discussion}\label{sec:discussion}

The values of the parameters constrained with our global fitting are listed
in Table~\ref{tab:result}. For comparison, the best fits of different datasets
with $n_\mathrm{t}$ as a free parameter and 
$n_\mathrm{t} = -r_\mathrm{0.05}/8$ following the consistency relation are all
listed in Table~\ref{tab:result}.

\begin{figure*}[htbp]
\begin{center}
\includegraphics[width=4.2in]{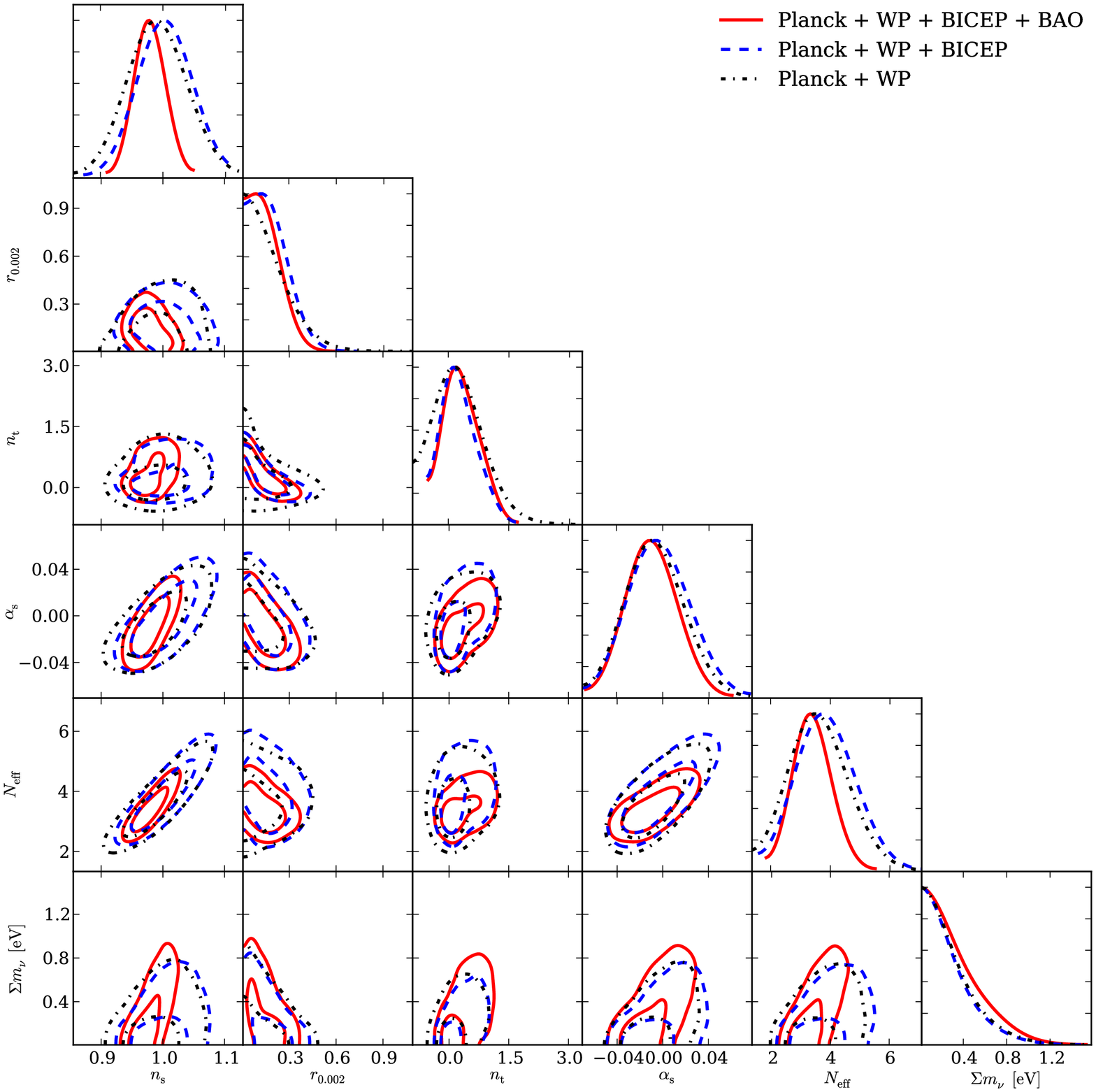}\\
\includegraphics[width=4.2in]{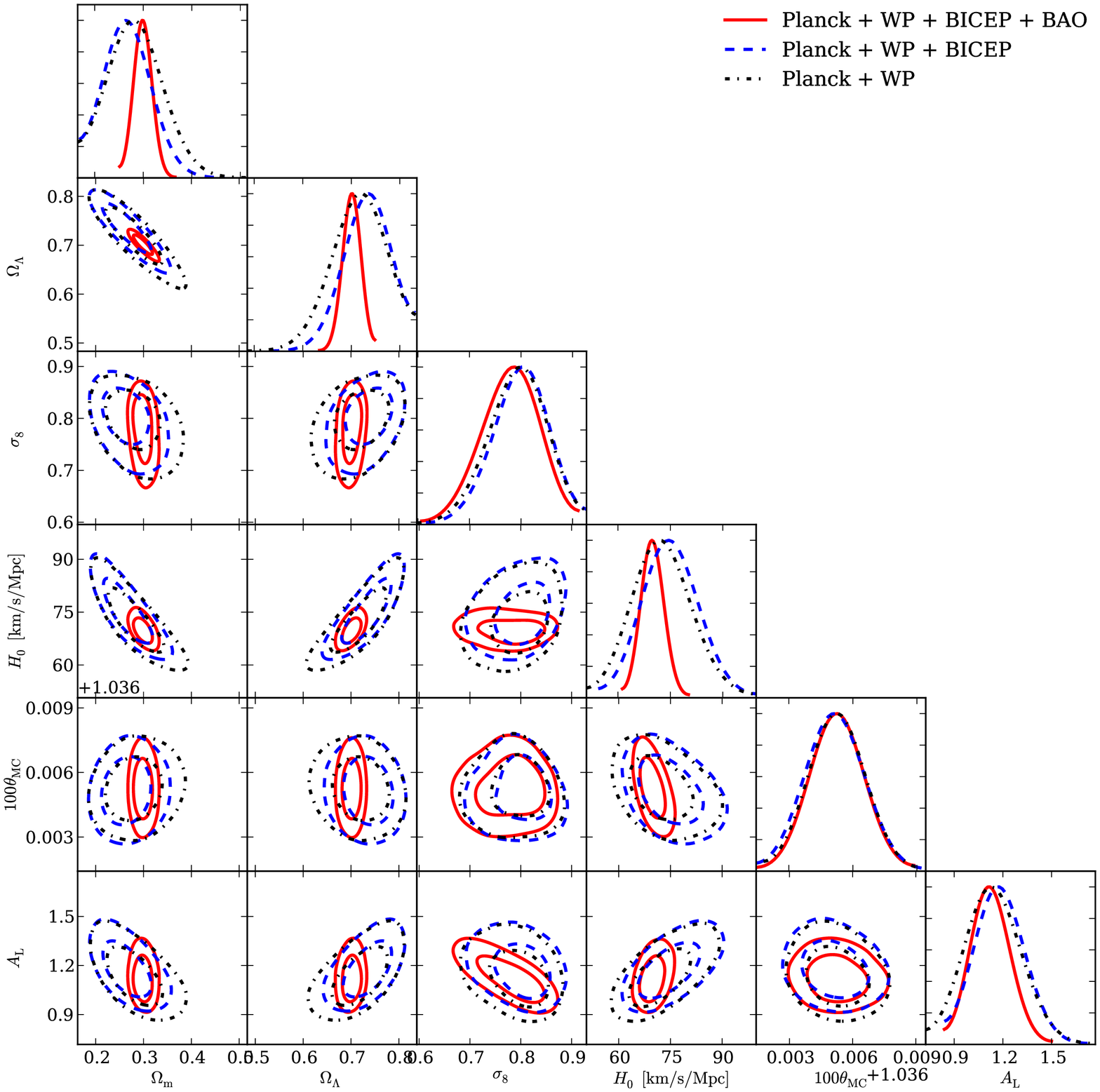}
\caption{
    The joint 1d and 2d probability distribution of cosmological 
    parameters.  The inner and outer contours
    represent the $68\%$ and the $95\%$ confidence levels respectively.
    Top: the primordial power spectrum
    parameters and the neutrino parameters 
    $N_{\mathrm{eff}}$ and $\Sigma \mathrm{m_{\nu}}$, 
    Bottom: other cosmological parameters.
    The consistency relation is not imposed.
}
\label{fig:tri_pw}
\end{center}
\end{figure*}

We plot the 2-dimensional contours and 1-dimensional probability distribution of
cosmological parameters with different data sets in Fig.\ref{fig:tri_pw}. 
The Planck+WMAP constraints
are plotted in black dash-dot lines, the constraint with additional BICEP data are plotted
with blue dash lines, and the constraint also with BAO are plotted in red solid lines. 
Here we do not impose the consistency relation, and $n_t$ is taken as a free parameter.
From these plots, we find that with the inclusion of the neutrino parameters, 
there is no significant conflict between the result of including and excluding
the BICEP2 data set, the allowed parameter range or region overlap with each other in these
two cases. The constraints also become tighter with the additional BAO datasets included.
With only the Planck and WMAP9 datasets, 
$r_\mathrm{0.002}<0.16(0.36)$ with $68\%(95\%)$ marginalized 
limits. The constraints are different from that reported by
\citet{2013arXiv1303.5076P}, since some extra free parameters 
are included in our global fitting, such as $n_\mathrm{t}$, 
$\alpha_\mathrm{s}$, $N_\mathrm{eff}$ and $\Sigma m_\mathrm{\mu}$.
Now, with the additional of the BICEP2 data, we find
$r_\mathrm{0.002} = 0.16_{-0.12}^{+0.06}$, and $r_\mathrm{0.002} = 0.14_{-0.11}^{+0.05}$ 
with both the BICEP2 and the BAO datasets included. The results are all 
consistent with each other. 

In the above we have taken $n_t$ as a free parameter without imposing the consistency relation.
If we do impose the consistency relation in our fitting, the 
$68\%$ marginalized limits  are  
$r_\mathrm{0.002} < 0.24$ with only Planck and WMAP9 datasets; 
$r_\mathrm{0.002} = 0.24_{-0.07}^{+0.05}$ with BICEP2 data included; and 
$r_\mathrm{0.002} = 0.22_{-0.06}^{+0.04}$ with both BICEP2 and BAO datasets 
included. The results are all consistent with Ref.\cite{2014arXiv1403.3985B}.

These results show  that the tension between the BICEP and Planck data is 
removed by including 
$n_\mathrm{t}$, $\alpha_\mathrm{s}$ and the neutrino parameters as free parameters
in the global fitting. Below we investigate which parameters are responsible for this.

\clearpage

\subsection{$n_\mathrm{t}$ is not the key parameter}\label{sec:nt}

As discussed in Sect.~\ref{sec:th_analysis}, $n_\mathrm{t}$ has 
significant effect on the BB power spectrum, but not on the TT or the matter power 
spectrum. The results of our global fitting with different data sets also indicate that
the constraints on $n_\mathrm{t}$ become better with the inclusdion of BICEP2 data set, 
but the BAO data set does not help to improve the constraint on it.

\begin{figure*}[htbp]
\begin{center}
\includegraphics[width=3.in]{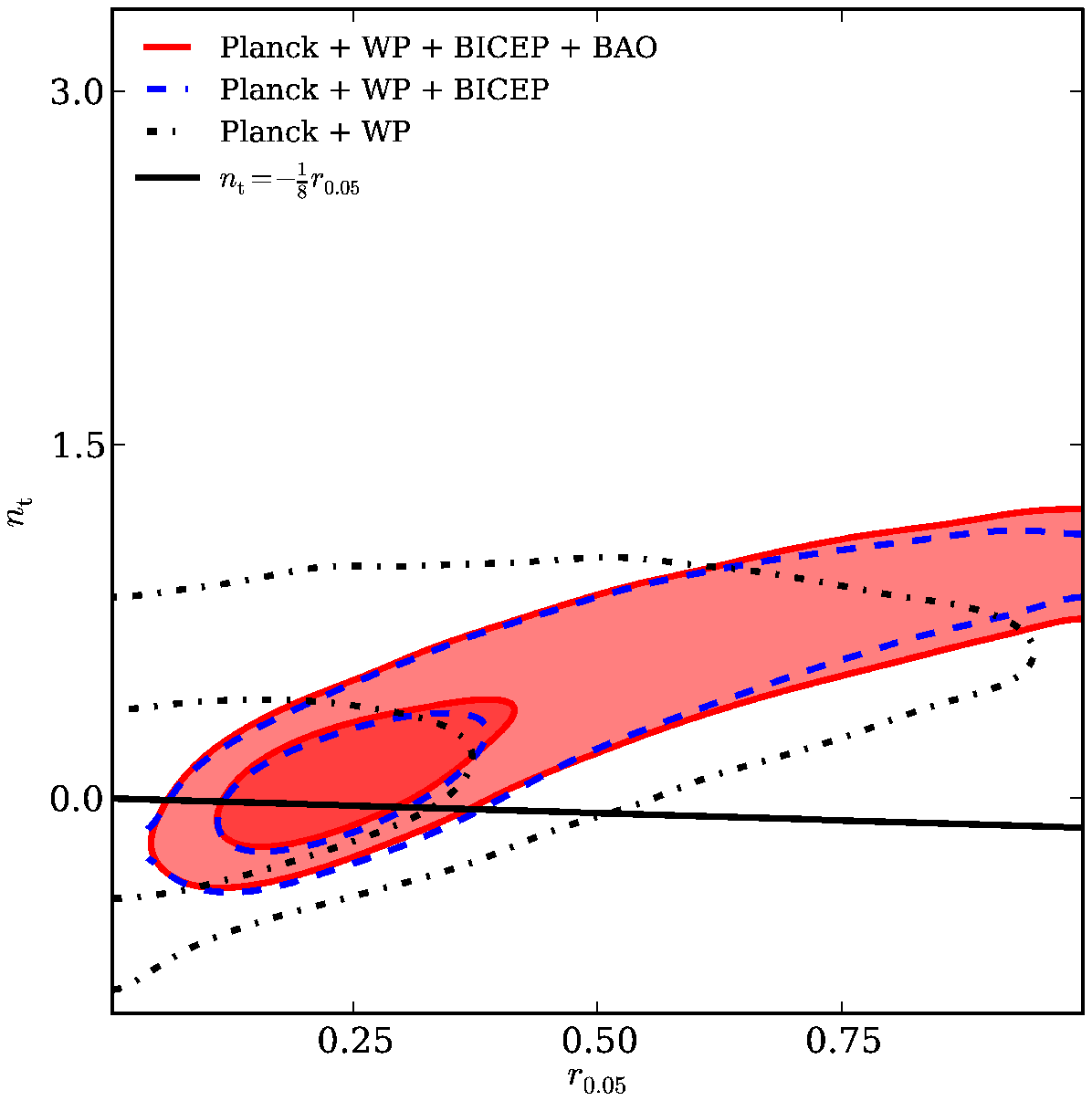}
\includegraphics[width=3.in]{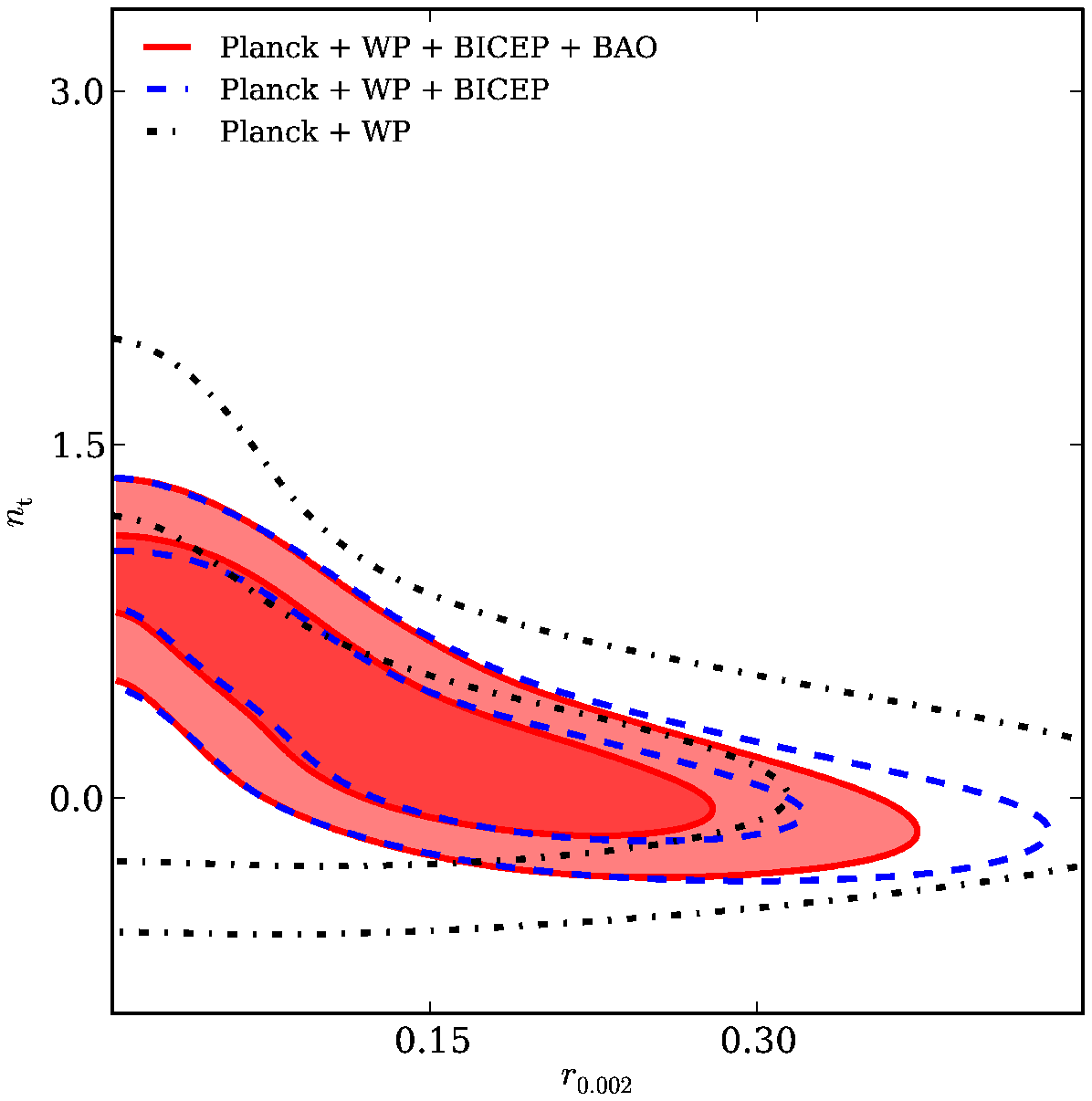}
\caption{
    Left: The joint probability of $n_\mathrm{t}$ and $r_\mathrm{0.05}$, 
the black solid line represents the single-field inflation 
    consistency relation $n_\mathrm{t} = -r_\mathrm{0.05} /8$.
    Right: The joint probability of $n_\mathrm{t}$ and $r_\mathrm{0.002}$.
The inner and outer contours represent the $68\%$ and $95\%$ confidence levels respectively. 
} 
\label{fig:nt_r}
\end{center}
\end{figure*}


The left panel of Figure~\ref{fig:nt_r} shows the joint probability of $n_\mathrm{t}$ and 
$r_\mathrm{0.05}$, when the consistency relation is not imposed. For reference, we  
also plot the consistence relation with the black solid line 
in the same figure. The consistency relation also forces 
$n_\mathrm{t}$ to be negative, so the tensor spectrum has a red tilt. From the figure
we see that the global fitting results are still consistent with the consistency relation.
Although the result favors a blue tilt in the tensor spectrum slightly, it is not
as significant as reported in Ref.\cite{2014arXiv1403.5732G}. 

In the right panel of Figure~\ref{fig:nt_r}, we show the contours for $r_{0.002}$ 
and $n_\mathrm{t}$, note here $n_\mathrm{t}$ is measured around $k=0.05 h/\Mpc$. 
If $r_{0.002}<0.11$\cite{2013arXiv1303.5076P}, it 
would be easy to get a larger blue tilt tensor power spectrum.  
However, when the consistency relation is imposed,
it forces $n_\mathrm{t}$ to a small negative value, and yields a large 
$r_{0.002}$. In our global fitting, the constraints with flat prior result in a slightly 
smaller $r_{0.002} = 0.14^{+0.05}_{-0.11}$, while with the consistency
relation imposed, $r_{0.002} = 0.22^{+0.05}_{-0.06}$, which is consistent with 
the results reported in Ref.\cite{2014arXiv1403.3985B}.

We see the results obtained with the different data sets 
are generally consistent with each other, either 
with or without the consistency relation imposed. Including $n_\mathrm{t}$ 
as a free parameter is not a necessary condition for solving the tension,
but the value of $r_\mathrm{0.002}$ is correlated with the prior of 
$n_\mathrm{t}$.

\subsection{$\alpha_\mathrm{s}$ helps little}\label{sec:alphas}

In the paper of \citet{2014arXiv1403.3985B}, $\alpha_\mathrm{s}$ is introduced to 
reduce the tension in $r_{0.002}$. According to their analysis, a negative 
$\alpha_\mathrm{s} \sim -0.022$ is needed for suppressing the scalar power spectrum.
In our fitting, the 
$\alpha_\mathrm{s}$ is constrained to 
$\alpha_\mathrm{s} = -0.0054_{-0.0217}^{+0.0162} (68\%)$ with Planck and WMAP9 data;
$\alpha_\mathrm{s} = -0.0011_{-0.0241}^{+0.0187} (68\%)$ with BICEP2 data included;
and $\alpha_\mathrm{s} = -0.0086_{-0.0189}^{+0.0148} (68\%)$ with all the datasets
included. The values of $\alpha_\mathrm{s}$ for different datasets agree within 
error range, and also consistent with $\alpha_\mathrm{s} = 0$.
Such a small $\alpha_\mathrm{s}$ can not give enough suppression on 
scalar perturbation for solving the tension.

In the case of $n_\mathrm{t}$ following the consistency relation assumption, the
$\alpha_\mathrm{s}$ is constrained to 
$\alpha_\mathrm{s} = -0.0133_{-0.0162}^{+0.0129} (68\%)$ with all the datasets
included. The non-zero results indicate that the $\alpha_\mathrm{s}$ is still helpful 
for suppressing the scalar power spectrum and alleviating the $r_\mathrm{0.002}$
tension, at least when the consistency relation is imposed. 

However,  the large value of $\alpha_\mathrm{s}\sim-0.02$ leads to 
small value of e-folds number in 
slow roll inflation, which is unacceptable in our 
universe \cite{2014arXiv1403.4596C,2006JCAP...09..010E},. In this case,
$\alpha_\mathrm{s}$ is not a good choice for solving the problem of tension between
Planck and BICEP2.

\subsection{The neutrinos helps much}\label{sec:neu}

\begin{figure}[htbp]
\begin{center}
\includegraphics[width=3.in]{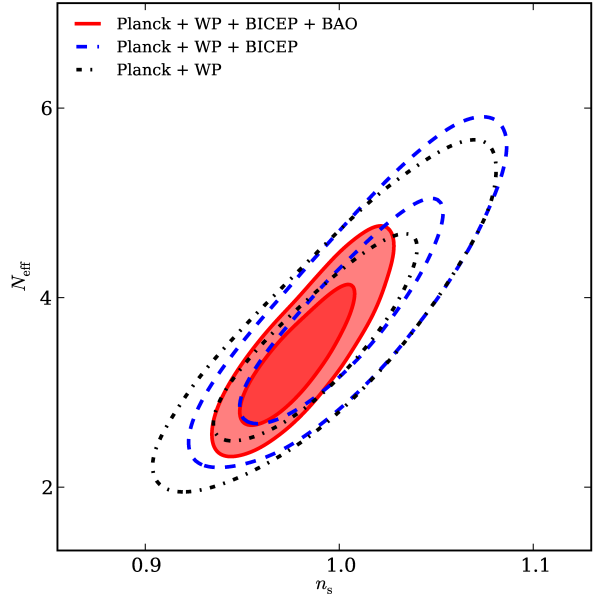}
\caption{
    This plot shows the joint probability of $N_\mathrm{eff}$ and $n_\mathrm{s}$
    constrained from our MCMC global fitting, by using different datasets, 
    without considering the consistency relation. The inner and outer contours
    represent the $68\%$ and $95\%$ confidence levels respectively. 
} 
\label{fig:nnu_ns}
\end{center}
\end{figure}

The neutrinos mainly affect the scalar and tensor power spectrum on small scales.
According to the analysis in \citet{2004PhRvD..69h3002B,
2013PhRvD..87h3008H, 2013neco.book.....L}, $N_{\mathrm{eff}}$ mainly affects the 
scales of BAO peaks, which are out of the scale range of BICEP2 data.  
So the constraint on neutrinos comes mainly from the Planck data sets and BAO data sets. 
As shown in the Figure~\ref{fig:tri_pw}, with neutrino parameters in the fit, 
the contours do not change much when the BICEP2 data is added. 
Because the neutrinos are still relativistic at the epoch of Recombination, 
$\Sigma \mathrm{m_{\nu}}$ only has a small effect on the primary power spectrum and 
it is hard to be constrained. We find that, $\Sigma \mathrm{m_{\nu}}$ is constrained
to be $< 0.29\mathrm{ev}$ with the Planck and the WMAP9 data; $< 0.27\mathrm{eV}$ with
the BICEP2 data included; and $< 0.37\mathrm{eV}$ with both the BICEP2 and the BAO 
data included. 

The $N_\mathrm{eff}$ is a more interesting parameter in this case. It is 
constrained to be $N_\mathrm{eff} = 3.66^{+0.61}_{-0.92}$
with the Planck and the WMAP9 data only; $N_\mathrm{eff} = 3.93^{+0.66}_{-0.93}$ 
with the BICEP2 data added; and $N_\mathrm{eff} = 3.43^{+0.39}_{-0.58}$ with
the BICEP2 and the BAO datasets included.
We get larger $N_{\mathrm{eff}}$ than that in the Standard Model. The result is 
consistent with recent CMB measurement \cite{2011ApJS..192...18K, 
2011ApJ...739...52D, 2011ApJ...743...28K, 2011PhRvD..84l3008A, 
2013ApJS..208...19H, 2013PhRvD..87h3008H, 2013arXiv1303.5076P}.  
Such a large $N_\mathrm{eff}$ is expected for solving
the $r_\mathrm{0.002}$ tension. Because of the suppression of the large 
$N_\mathrm{eff}$ on small scales,
a large $n_\mathrm{s}$ becomes acceptable, and the large $n_\mathrm{s}$ can help 
solving the tension problem on large scale. 
Such a degeneration can be found through the 
2d contours of $N_\mathrm{eff}$ and $n_\mathrm{s}$ in Figure~\ref{fig:nnu_ns}.
Without considering BAO and BICEP2 data, 
$n_\mathrm{s} = 0.99_{-0.04}^{+0.03}$
which is larger than $1$ within $68\%$ marginalized confidence interval. 
With the BICEP2 data added, the $n_\mathrm{s}$ is constrained to be
$1.01^{+0.03}_{-0.04}$.
By including the BAO and BICEP2 data,
$n_\mathrm{s} = 0.98_{-0.02}^{+0.02}$
which is still consistent with $n_\mathrm{s} < 1$.  

The large $N_{\mathrm{eff}}$ could be explained by including extra neutrinos such as the
sterile neutrinos, neutrino/anti-neutrino asymmetry and/or any other light 
relics in the universe. In the case of sterile neutrinos, 
the related parameters, $m_{\nu,\,\mathrm{sterile}}^{\mathrm{eff}}$ and 
$N_\mathrm{eff}$ are constrained to 
$m_{\nu,\,\mathrm{sterile}}^{\mathrm{eff}} < 0.79$, $N_\mathrm{eff} < 4.30$ with
only the Planck and the WMAP9 datasets; 
$m_{\nu,\,\mathrm{sterile}}^{\mathrm{eff}} < 0.75$, 
$N_\mathrm{eff} = 4.19^{+0.36}_{-1.08}$
with the BICEP2 data added;
and $m_{\nu,\,\mathrm{sterile}}^{\mathrm{eff}} = 0.53^{+0.21}_{-0.42}$,
$N_\mathrm{eff} < 4.05$ with
the BICEP2 and the BAO datasets both included. 
And in such case, $r_\mathrm{0.002}$ is constrained to 
$r_\mathrm{0.002} < 0.23$ with the Planck and the BICEP2 datasets;
$r_\mathrm{0.002} = 0.18^{+0.08}_{-0.11}$ with the BICEP2 datasets included;
and $r_\mathrm{0.002} = 0.19^{+0.08}_{-0.09}$ with the BICEP2 and the BAO datasets
both included. The consistent constraints on $r_\mathrm{0.002}$ indicate 
the alleviation of the tension between different datasets. 

Using sterile neutrinos to alleviate the tension between the Planck data set and 
other data set has also been discussed in \citet{2014arXiv1403.7028Z, 2014arXiv1403.8049D}.
Their conclusion are in agreement with  ours. 
With the complete exploration of the parameter space, we also find that
including neutrino parameters plays an important role in solving the tension.


\section{conclusion}\label{sec:conclusion}

In this paper, we explore various ways to alleviate apparent tension between 
the constraints on the inflationary tensor-to-scalar ratio $r_\mathrm{0.002}$
obtained from the BICEP2 data and the Planck data. The fittings are performed with the
Planck CMB temperature data\cite{2013arXiv1303.5076P} and 
the WMAP 9 year CMB polarization data\cite{2013ApJS..208...19H, 2013ApJS..208...20B},
with/out the newly published BICEP2 CMB B-mode data. we also use the BAO data from 
SDSS DR9\cite{2014MNRAS.441...24A}, SDSS DR7\cite{2012MNRAS.427.2132P} and 
6dF\cite{2011MNRAS.416.3017B}, 
to help breaking some parameter degeneracy and improve the precision of the model..
By setting $\alpha_\mathrm{s}$, $n_\mathrm{t}$ and neutrino parameters as 
free parameters, the resulting constraints on $r_\mathrm{0.002}$ from different data sets
 are found to be consistent with each other, 

With all the datasets included, we obtain 
marginalized $68\%$ bounds on some interested parameters as follows:
\begin{eqnarray} 
    r_\mathrm{0.002} &=&  0.14_{-0.11}^{+0.05} , \\ 
        n_\mathrm{s} &=&  0.98_{-0.02}^{+0.02} , \\ 
   \alpha_\mathrm{s} &=& -0.0086_{-0.0189}^{+0.0148} , \\
        n_\mathrm{t} &=&  0.35_{-0.47}^{+0.28} .
\label{eq:final}
\end{eqnarray}
The value of $r_\mathrm{0.002}$ obtained in this work 
is smaller than that reported by the BICEP2 team, 
due to its dependence on $n_\mathrm{t}$, which is constrained to be 
positive(blue tensor tilt), but a flat or even red tilt is still consistent with the data.  
Further more, the results do not deviate from the consistency 
relation, even if we ignore the relation in the fitting.
Because the consistency relation restricts $n_\mathrm{t}$ to a lower 
value, it breaks the degeneracy between $n_\mathrm{t}$ and $r_{0.002}$. 
By applying this relation as a prior in the fitting, a tighter 
constraint on $r_{0.002}$ is obtained, $r_{0.002}=0.22_{-0.06}^{+0.05}$.

Although the tension is alleviated by including $\alpha_\mathrm{s}$, $n_\mathrm{t}$ 
and neutrino parameters as free parameters, we find that $\alpha_\mathrm{s}$ and
$n_\mathrm{t}$ are not the key parameters.
The scalar running $\alpha_\mathrm{s}$ is still consistent 
with $0$, this indicates that including $\alpha_\mathrm{s}$ may not be the best 
choice for solving the $r_{0.002}$ tension;
The results are consistent with different data sets, with or without 
$n_\mathrm{t}$ as a free parameter, which indicates that $n_\mathrm{t}$ 
is not necessary for solving the $r_\mathrm{0.002}$ tension problem.
Finally, the effective number of neutrinos $N_{\mathrm{eff}}$, constrained to 
$3.43_{-0.58}^{+0.39}$, appears to be the most important parameter for this problem.

We also check our result with the sterile neutrinos. By including all the data sets, 
$N_{\mathrm{eff}}$ is constrained to be $<4.05$, 
$m_{\nu,\,\mathrm{sterile}}^{\mathrm{eff}}$ is constrained to be 
$0.53^{+0.21}_{-0.42}$, and in this case, $r_\mathrm{0.002}$ is constrained to be
$0.19^{+0.08}_{-0.09}$.


\section*{Acknowledgements}
We thank  Antony Lewis for kindly providing us the beta version of  CosmoMC code 
for testing. Our MCMC computation was performed on the Laohu cluster in NAOC and 
on the GPC supercomputer at the SciNet HPC Consortium.
This work is supported by the Ministry of Science and Technology 863 project
grant 2012AA121701, the NSFC grant 11073024, 11103027, and the CAS Knowledge
Innovation grant KJCX2-EW-W01.

\bibliography{bicep2}
\bibliographystyle{apsrev}
\end{document}